\documentclass{article}
\usepackage{hiph-art}
\volnumber{19} \issuenumber{1} \edyear{2004}                             
\frompage{000} \topage{000}                                              
\recrevdate{31 October 2005}                                              
\def\rightmark{Zipf's Law in Fragmentation}
\def\leftmark{W. Bauer, S. Pratt, B. Alleman}
 
\title{Zipf's Law and the Universality Class of the Fragmentation Phase Transition} 
\authors{ 
{Wolfgang Bauer$^{1,2,a}$, Scott Pratt$^{1,b}$, and Brandon Alleman$^{3,c}$ %
\index{Bauer, W.} 
\index{Pratt, S.} 
\index{Alleman, B.} 
}\\[2.812mm]
{\normalsize
\hspace*{-8pt}$^1$ Department of Physics and Astronomy\\
Michigan State University\\
East Lansing, MI 48824-2320, USA\\[0.2ex] 
\hspace*{-8pt}$^2$ National Superconducting Cyclotron Laboratory\\
Michigan State University\\
East Lansing, MI 48824-1321, USA\\[0.2ex] 
\hspace*{-8pt}$^3$ Hope College\\ 
Holland, MI 49423, USA
}}
 
\abstract{We show that Zipf's Law for the largest clusters is not valid in an exact
sense at the critical point of the fragmentation
phase transition, contrary to previous claims.  Instead, the extracted distributions of
the largest clusters reflects the choice of universality class through the value of
the critical exponent $\tau$.}

 
\makeindex
\begin{document}
 
\maketitle

\section{Multifragmentation}

Nuclear multifragmentation offers a unique opportunity to study phase
transitions under extreme finite size constraints.  Throughout the last decade
models of this phase transition have been developed and tested against
experimental data.  In particular, recent scaling analysis of multifragment emission
has yielded important information on the character of the phase transition,
yielding values for the critical temperature and the values of the exponents
$\tau$ and $\sigma$ \cite{kleine,elliott}.

Event samples provided by multifragmentation are also a rich source for
statistical analysis.  As one example, we cite the observation that intermediate-mass fragment
multiplicity distributions as a function of the total transverse energy follow sub-Poissonian
distributions \cite{moretto}. Our group, however, was able to show that this binomial scaling
was exclusively a finite-size effect \cite{gharib,bauer98,bauer99}.  Furthermore, we were able to
show that the same patterns exist in word-length frequencies in the literature, where the system
size is given by the length of a paragraph \cite{bauer99a}.

In 1949, the linguist G. Zipf published \cite{zipf} another very interesting observation about many languages: if one ranks all words in a wide variety of texts in a given language, then the rank is inversely proportional to the frequency of occurrence
\begin{equation}\label{ZipfLaw}
	F(r) = c\,r^{-\lambda}
\end{equation}
(with $c$ = constant, $\lambda\approx 1$). The same distribution was found much earlier by Pareto in ranking companies by their income \cite{pareto}.  Similarly, ranking US cities by their population yields the same law.  Recently, Watanabe \cite{watanabe} observed in numerical simulations that percolation clusters at the critical threshold also obey Zipf's Law, and Ma \cite{ma,ma05} proposed to make use of this finding to detect a crossing of the critical threshold.  This motivates the current study.
 
\section{Cluster Sizes at the Critical Point}
 
In a scaling theory the cluster size distribution follows the scaling function
\begin{equation}
n(A,\epsilon) = a\,A^{-\tau}f(A^\sigma\cdot\epsilon)
\end{equation}
Here $\epsilon$ is the fractional deviation of the control parameter from the
critical value, $\epsilon = (p-p_c)/p_c$, $A$ is the cluster size (= number of
nucleons in the case of nuclear fragmentation), $f$ is the scaling function
with the boundary condition of $f(0)=1$, and $\sigma$ and $\tau$ are the
critical exponents that determine the universality class of the phase transition. 
The normalization constant $a$ can be obtained from the condition that all
nucleons have to belong to some cluster:
\begin{equation}
	\sum_{A=1}^{V} A\cdot n(A,\epsilon) = V,
\end{equation}
where $V$ is the event size (= total number of nucleons).

Let us first address scaling at the critical point,
\begin{equation}
	\epsilon = 0 \Rightarrow n(A,0) \equiv N(A) = a\,A^{-\tau} ,
\end{equation}
i.e. the cluster size distribution is a pure power law.

The probability $P_{1st}(A)$ that a cluster of size $A$ is the largest in a given
fragmentation event is the product of the probability that there is at least
1 cluster of size $A$ present, and that there are 0 clusters of size larger
than $A$ present in the same event.
\begin{equation}
	P_{1st}(A) = p_{\le 1}(A) \cdot p_0(>\hspace*{-0.1cm}A) 
	= [1-p_0(A)]\cdot p_0(>\hspace*{-0.1cm}A).
\end{equation}

Since the average yield of clusters of size $A$ is $N(A) = a\,A^{-\tau}$, we
obtain for the average yield of clusters of size bigger than $A$ the sum
\begin{equation}
	N(>\hspace*{-0.1cm}A) = \sum_{i=A+1}^{V} N(i) = \sum_{i=A+1}^{V} a\,i^{-\tau}
	= a\, \zeta(\tau, 1+A) - a\, \zeta(\tau, 1+V)
\end{equation}
where $\zeta$ is the generalized Riemann function, and we find for the normalization constant
\begin{equation}\label{norm}
	a = V / \sum_{A=1}^V A^{1-\tau} = V/H_V^{1-\tau}
\end{equation}
with $H_V^{1-\tau}$ is the $V^{th}$ harmonic number of order $1-\tau$.

For $V\gg A$, the probabilities to have $k$ clusters of size $A$ in a given event are Poissonian,$^{d}$
\begin{equation}
	p_k(i) = \frac{1}{n!} \langle N(i) \rangle^k e^{\langle N(i) \rangle}
\end{equation}

Thus, we finally obtain for the probability that a cluster of size $A$ is the largest in a
given fragmentation event
\begin{equation}
	P_{1st}(A) =  [1-p_0(A)]\cdot p_0(>\hspace*{-0.1cm}A)
	= [1-e^{-a\,A^{-\tau}}]\cdot e^{[a\, \zeta(\tau, 1+A) - a\, \zeta(\tau, 1+V)]},
\end{equation}
with $a$ given by equation \ref{norm}.

From this we obtain the exact expression for the average size of the largest cluster by summation over all $A$:
\begin{equation}
	\langle A_{1st} \rangle = \sum_{A=1}^{V} A\cdot P_{1st}(A)
\end{equation}

The probability that a cluster of size $A$ is the second biggest in a given event is the sum of two terms. The first term is the probability that two or more clusters of size $A$ are present times the probability that none bigger than $A$ is present.  And the second term is the probability that exactly one cluster of size larger than $A$ is present times the probability that at least one cluster of size $A$ is present.  This yields:
\begin{eqnarray}
	 P_{2nd} (A) &=& p_{ \ge 2} (A) \cdot p_0 ( > A) + p_{ \ge 1} (A) \cdot p_1 ( > A) \nonumber\\ 
  	&=& [1 - p_0 (A) - p_1 (A)] \cdot p_0 ( > A) + [1 - p_0 (A)] \cdot p_1 ( > A)
\end{eqnarray}
Obviously, this can be continued via recurrence relations to the third, fourth, ... $n$th biggest clusters. It is important to note that these exact analytical results only depend on only two parameters, the size of the fragmenting system, $V$, and the value of the critical exponent $\tau$.

\begin{figure}[htb]
          \insertplot{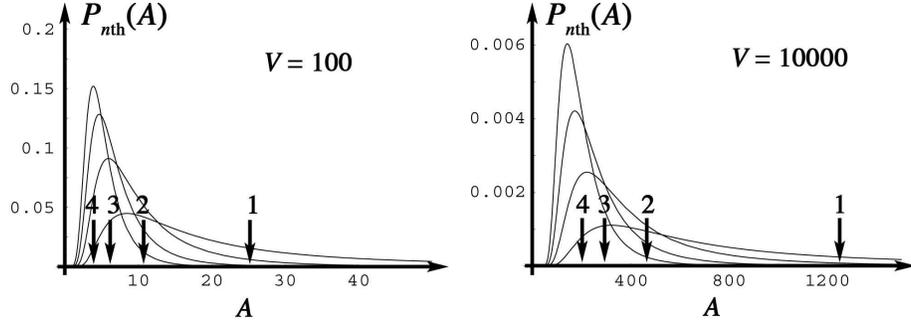 \epsfysize=4.2cm}
\vspace*{-0.8cm}
\caption[]{Probability distributions for the largest to fourth-largest clusters in a fragmentation event of
a system of size $V = 100$ (top) and $V = 10,000$ (bottom).
The arrows indicate the corresponding average values.}
\label{fig1}
\end{figure}

As an example, in figure~\ref{fig1} we show the probability distributions for the  largest through 
fourth-largest fragments as a function of mass number $A$ for two different system sizes, $V=100$ and $V=10,000$.  Also marked in this figure, by the vertical arrows, are the average values for the largest through fourth-largest fragment.  In both cases the value for $\tau$ was chosen as 2.18, that of a 3d percolation universality class. Next we investigate whether Zipf's Law, equation~\ref{ZipfLaw}, is satisfied.

\begin{figure}[htb]
\vspace*{-1.4cm}
                 \insertplot{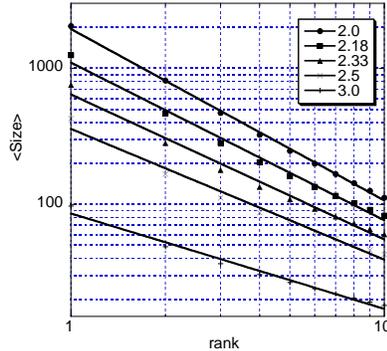}
\vspace*{-2.0cm}
\caption[]{Average size of the largest fragment of rank $r$ for a system
size of $V = 10,000$. The solid lines are power-law fits. The different sets of data were generated
with different values of the critical exponent $\tau$, as indicated in the inset.}
\label{fig2}
\end{figure}

In figure~\ref{fig2} we plot the average size of clusters of rank $r$ for 5 different values of the critical exponent $\tau$ in a system of size $V=10,000$ (plot symbols), together with their best respective power law fits of the form $\langle A_r \rangle \propto r^{-\lambda}$. It is clear that these data approximately follow power laws, but not {\it exactly}.  The power law exponents $\lambda$ numerically extracted are $1.26,1.16,1.07,0.97,0.7$ for $\tau= 2, 2.18,2.\overline{3},2.5,3$, respectively. One can see that in all cases our results are close to Zipf's Law. But there is a systematic dependence of $\lambda$ on the critical exponent $\tau$, which is something that should have been expected.

If one uses an expansion for our analytical results, one finds that the average cluster size as a function of rank $r$ follows a more general Zipf-Mandelbrot distribution \cite{mandel1,mandel2}
\begin{equation}\label{zipfmandelbrot}
	\langle A_{r{\rm th}}\rangle = c (r + k)^{-\lambda}
\end{equation}
where the offset $k$ is an additional constant that one has to introduce, and $\lambda$ is asymptotically approximated as a function of the critical exponent $\tau$
\begin{equation}\label{lambdaoftau}
	\lambda(\tau) = \frac{1}{\tau-1}
\end{equation}

\begin{figure}[htb]
                 \insertplot{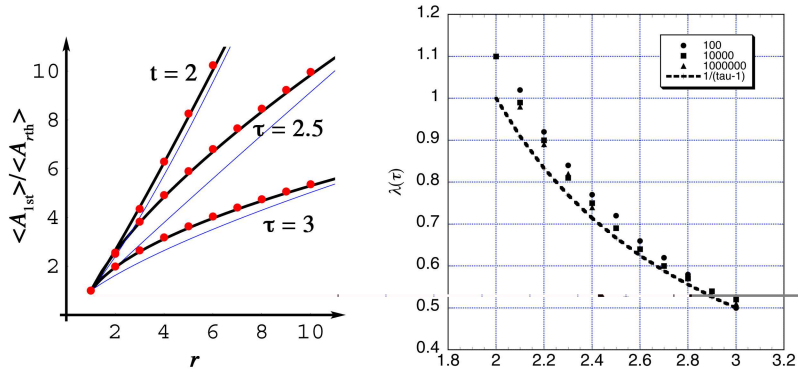  \epsfysize=5.2cm} 
\vspace*{-1.0cm}
\caption[]{Left: Ratio of the average size of the largest cluster to that of the $r$th largest. Dots are the analytical results, thin lines the best simple power law fits, thick lines the fits with the distribution of equation \ref{zipfmandelbrot}. Right: Effective exponent $\lambda$ of the Zipf-Mandelbrot distribution fit to the exact fragmentation cluster size distribution as a function of the critical exponent $\tau$ for three different system size, $V=100$, $V=10,000$, and $V=1,000,000$. Dashed line: asymptotic expansion of 
equation~\ref{lambdaoftau}}
\label{fig3}
\end{figure}

One obtains almost perfect fits to the average cluster sizes as function of their rank by using equation~\ref{zipfmandelbrot}. This is particularly evident if one plots the ratio of the average size of the largest cluster to that of the $r$th largest cluster, see left side of figure~\ref{fig3}.  The resulting value of the exponent $\lambda$ as a function of the critical exponent $\tau$ are shown on the right side of  figure~\ref{fig3}.  The dashed line is the approximation of equation~\ref{lambdaoftau}.

\section{Conclusions}\label{concl}
We have shown that the distribution of largest clusters does not exactly
follow the functional form implied by Zipf's Law, but instead a more general distribution of the Zipf-Mandelbrot kind, and with a power-law exponent that is not unity, but varies as a function of the critical exponent $\tau$.  Thus the observation of frequency distributions close
to the expectation of Zipf's Law does not yield additional physics insight
and must be considered coincidental. 
 
\section*{Acknowledgments}
This research was supported by the US National Science Foundation under
grants PHY-0245009 (W. Bauer) and PHY-0243709 (B. Alleman), as well as
the US Department of Energy under grant number DE-FG02-03ER41259 (S. Pratt).
 
\section*{Notes} 
\begin{notes}
\item[a] email: bauer@pa.msu.edu
\item[b] email: pratts@pa.msu.edu
\item[c] REU summer student at Michigan State University, summer 2005
\item[d] We have also conducted studies in the percolation model, where the independent Poissonian emission condition does not strictly hold, with similar results.
\item[e] After this project was completed, we became aware through private discussion
that X. Campi and H. Krivine have arrived at somewhat similar conclusions independently,
and in a different way \cite{Cam05}.
\end{notes}

\vfill\eject
\end{document}